\documentclass[10pt,conference,twocolumn]{IEEEtran}
\usepackage{mathrsfs}
\usepackage{amsfonts}
\usepackage{mdwmath}
\usepackage{mdwtab}
\usepackage{cite}
\usepackage{graphicx}
\usepackage[cmex10]{amsmath}
\usepackage{array}
\usepackage{algorithmic}
\usepackage{algorithm}
\usepackage{blindtext}
\usepackage[english]{babel}
\usepackage{mathrsfs}

\renewcommand{\baselinestretch}{1}
\renewcommand{\figurename}{Fig.}
\newcounter{RomanNumber}
\newcommand{\MyRoman}[1]{\setcounter{RomanNumber}{#1}\Roman{RomanNumber}}
\begin{document}
\renewcommand{\figurename}{Fig.}  
\title{A Tensor Completion Approach for Efficient and Robust Fingerprint-based Indoor Localization}
\author{\IEEEauthorblockN{Yu Zhang\IEEEauthorrefmark{1}, Xiao-Yang Liu\IEEEauthorrefmark{2}}
\IEEEauthorblockA{\IEEEauthorrefmark{1}Shanghai Institute of Measurement and Testing Technology, China}
\IEEEauthorblockA{\IEEEauthorrefmark{2}Shanghai Jiao Tong University, China\\}
\IEEEauthorblockA{\IEEEauthorrefmark{1}zhang\_yu@simt.com.cn, \IEEEauthorrefmark{2}yanglet@sjtu.edu.cn}
}
\IEEEtitleabstractindextext{%
\begin{abstract}
The localization technology is important for the development of indoor location-based services (LBS). The radio frequency (RF) fingerprint-based localization is one of the most promising approaches. However, it is challenging to apply this localization to real-world environments since it is time-consuming and labor-intensive to construct a fingerprint database as a prior for localization. Another challenge is that the presence of anomaly readings in the fingerprints reduces the localization accuracy. To address these two challenges, we propose an efficient and robust indoor localization approach. First, we model the fingerprint database as a 3-D tensor, which represents the relationships between fingerprints, locations and indices of access points. Second, we introduce a tensor decomposition model for robust fingerprint data recovery, which decomposes a partial observation tensor as the superposition of a low-rank tensor and a spare anomaly tensor. Third, we exploit the alternating direction method of multipliers (ADMM) to solve the convex optimization problem of tensor-nuclear-norm completion for the anomaly case. Finally, we verify the proposed approach on a ground truth data set collected in an office building with size 80\,m $\times$ 20\,m. Experiment results show that to achieve a same error rate 4$\%$, the sampling rate of our approach is only 10$\%$, while it is 60$\%$ for the state-of-the-art approach. Moreover, the proposed approach leads to a more accurate localization (nearly 20$\%$, 0.6\,m improvement) over the compared approach.
\end{abstract}
%
}
\maketitle
\IEEEdisplaynontitleabstractindextext

\section{Introduction}
Indoor location-based services (LBS) have received more and more attention with applications to many fields, such as navigation, health care and location-based security. The indoor localization is a supporting technology for LBS. The RF fingerprint-based approach is regarded as a promising solution, which can exploit ubiquitous infrastructure of WiFi access points and achieve higher localization accuracy \cite{ficco2014calibrating} \cite{6866898} \cite{redzic2014seamloc}. It consists of two phases: a training phase (site survey) and an operating phase (online location). In the first phase, RSS fingerprints are collected from different access points (APs) at reference points (RPs) and constructs a fingerprint database (radio map), in which fingerprints are related with the corresponding RPs. In the second phase, the real-time RSS samples measured by user's mobile device are compared with the stored fingerprint database to estimate the user's current location.

However, the major challenge for the RF fingerprint-based approach comes from the time-consuming and labor-intensive site survey process, which collects RSS samples at dense RPs for constructing the fingerprint database. It will be a very exhaustive process and lead to a significant workload. Furthermore, the RSS fingerprints are usually corrupted by ubiquitous anomaly readings, which may come from the people moving, measuring device itself or the dynamic surroundings. As a result, it reduces the accuracy of the fingerprint data and then increase the localization error. Therefore, it is necessary to distinguish fingerprints from anomaly readings in a robust and accurate way.

\begin{figure}[!hbt]
\centering
\includegraphics[width=3in]{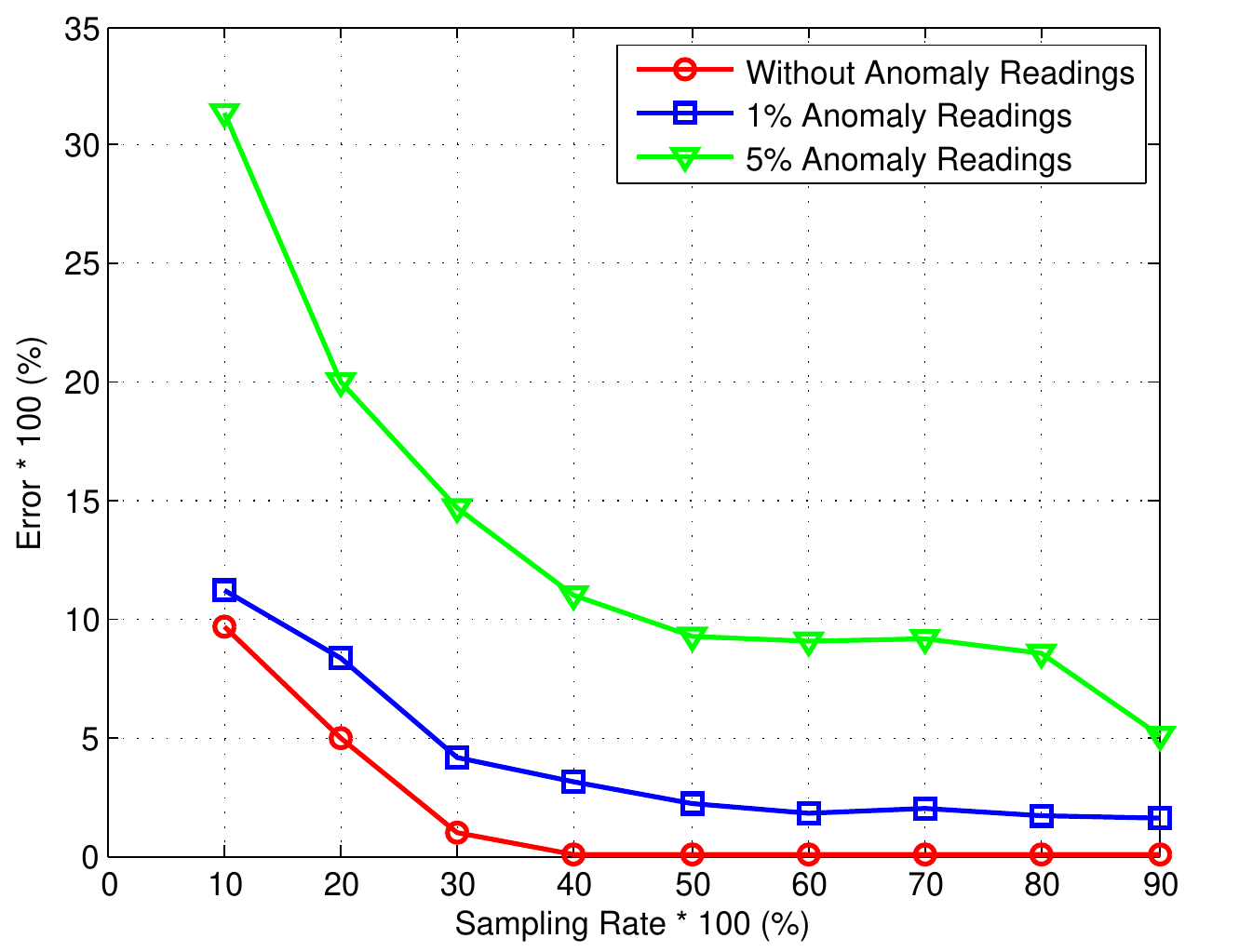}
\caption{Tensor completion with anomaly readings.}
\label{fig1}
\end{figure}

Some research works have been dedicated to relieve the site survey burden. The radio propagation model can be exploited to construct the fingerprint database from a small number of measurements \cite{xiang2005hidden}. However, it needs to overcome the influence produced by multipath fading and unpredictable nature of radio channel. The interpolation approaches are used to construct a complete fingerprint database, which usually estimate the missing RSS data by the interpolation of the measurements at local adjacent RPs \cite{kuo2011discriminant}. They still demand collecting a lot of fingerprint data for obtaining better localization accuracy. Recent approaches include the matrix completion \cite{hu2013efficient} and tensor completion \cite{liu2015fine}, which recover the fingerprint data by exploiting the inherent spatial correlation of RSS measurements. However, the recovery accuracy may be reduced because of the presence of anomaly readings, which decrease the localization performance.
To explain the impact of the anomaly readings to data recovery accuracy in tensor completion, we first verify that the tensor completion experiences performance degradation with the presence of a small portion of anomalies. Fig. \ref{fig1} is a simple example, a 100 $\times$ 100 $\times$ 30 low-rank tensor taking values in [0,100], can be exactly recovered using 40$\%$ uniformly random selected samples. If we add anomalies with two different ratios 1$\%$ and 5$\%$ into the constructed tensor, the recovery errors increase significantly. The essential reason is that a portion of anomaly readings break down the low-rank structure of tensor.

Therefore, in this paper, we aim to reduce the efforts during the site survey, and at the same time, improve the recovery accuracy of fingerprint data. Firstly, we exploit the notion of tensor rank proposed in \cite{kilmer2013third} to model the RF fingerprint data as a 3-D tensor. Secondly, we propose a tensor decomposition model in the anomaly case for robust fingerprint data recovery. The model decomposes a corrupted observation tensor as the superposition of a low-rank tensor and a spare anomaly tensor. Thirdly, we formulate the robust fingerprint data recovery problem as an optimization problem and exploit the ADMM algorithm for fingerprints completion. Finally, evaluations on a real fingerprint data set show that our scheme outperforms existing approach.

Our contributions are as follows:

$\bullet$ To the best of our knowledge, this is the first work to study the anomaly tensor completion approach for fingerprint database construction.

$\bullet$ We exploit the ADMM algorithm to solve the proposed tensor completion problem.

$\bullet$ The performance of the proposed tensor completion approach is evaluated based on the ground truth collected in real experiment field. Under the sampling rate 20$\%$, our approach achieves less than 5$\%$ recovery error. The localization accuracy has the improvement of 20$\%$ compared with the existing approach.

The remainder of this paper is organized as follows. In Section \MyRoman{2}, we present the related works. Section \MyRoman{3} shows the problem formulation. Robust tensor completion is given in Section \MyRoman{4}. The performance is evaluated through implementations in Section \MyRoman{5}. Finally, Section \MyRoman{6} concludes the paper.
\section{Related Work}
Constructing fingerprint database is an important step for fingerprint-based localization, which has a great impact on the accuracy of localization. In order to ensure accuracy, a complete database is needed for localization. Some approaches are proposed for achieving complete fingerprint data while reducing the site survey efforts. These approaches fall into two categories: correlation-aware approach and sparsity-aware approach.

The correlation-aware approach exploits the fact that the fingerprints at nearby RPs are spatially correlated. For example, the radio propagation model can construct the fingerprint database by inserting some virtual data from the real measurements \cite{xiang2005hidden}, which is not accurate in complex indoor environment because of signal reflection and diffraction. So the interpolation approaches are used to construct a complete fingerprint database \cite{kuo2011discriminant} \cite{minimizing2003}, which usually estimate the missing RSS data by the interpolation of the measurements at local adjacent RPs. \cite{li2005method} presented a krigging interpolation method that can achieve more accurate estimation and reduce the workload. Such schemes outperform the radio propagation model approach, but they demand collecting a lot of fingerprint data for obtaining better localization accuracy.

Sparsity-aware approach considers the inherent spatial correlation of RSS measurements. It exploits the compressive sensing (CS) to construct complete fingerprint database from partial observation fingerprints \cite{feng2012received}.\cite{zhang2013using} assumed that RSS matrix for each AP was low-rank. \cite{hu2013efficient} further assumed that the low-rank matrix was smooth and exploits this property to construct fingerprint database. \cite{milioris2015building} exploited the spatial correlation structure of the fingerprints and use the framework of matrix completion to build complete training maps from a small number of random sample fingerprints. Recently, as extension of matrix completion and a strong tool, tensor completion is applied to many domains for data completion from limited measurements. \cite{liu2015fine} proposed an adaptive sampling method of tensor fibers based on which, the fingerprint data is constructed using tensor completion. It is the most related work to our approach. Through proper sampling, these approaches can exactly recover the data from a small number of entries via low-rank optimization. However, fingerprints usually suffer from anomaly readings, the influence of the anomaly readings to data recovery accuracy is not considered in it.
All above approaches mainly aim at reducing the site survey efforts. However, the anomaly readings are common and ubiquitous and effect the recovery accuracy of data. Our work is to further improve the recovery accuracy by distinguish fingerprint data from anomaly readings while reducing calibration efforts. Further more, the localization accuracy can be improved using the constructed fingerprint database.
\section{Problem Formulation}
In this section, we first briefly describe our localization system architecture. Then we present the notations used throughout the paper, the algebraic framework and preliminary results for third-order tensors as proposed in \cite{liu2015fine} \cite{kilmer2013third} \cite{zhang2014novel}. Finally, according to the fundamental theorems of tensor completion, we formulate the system model and state the problem for efficient and robust fingerprint-based indoor localization.
\begin{figure}[H]
\centering
\includegraphics[width=3in]{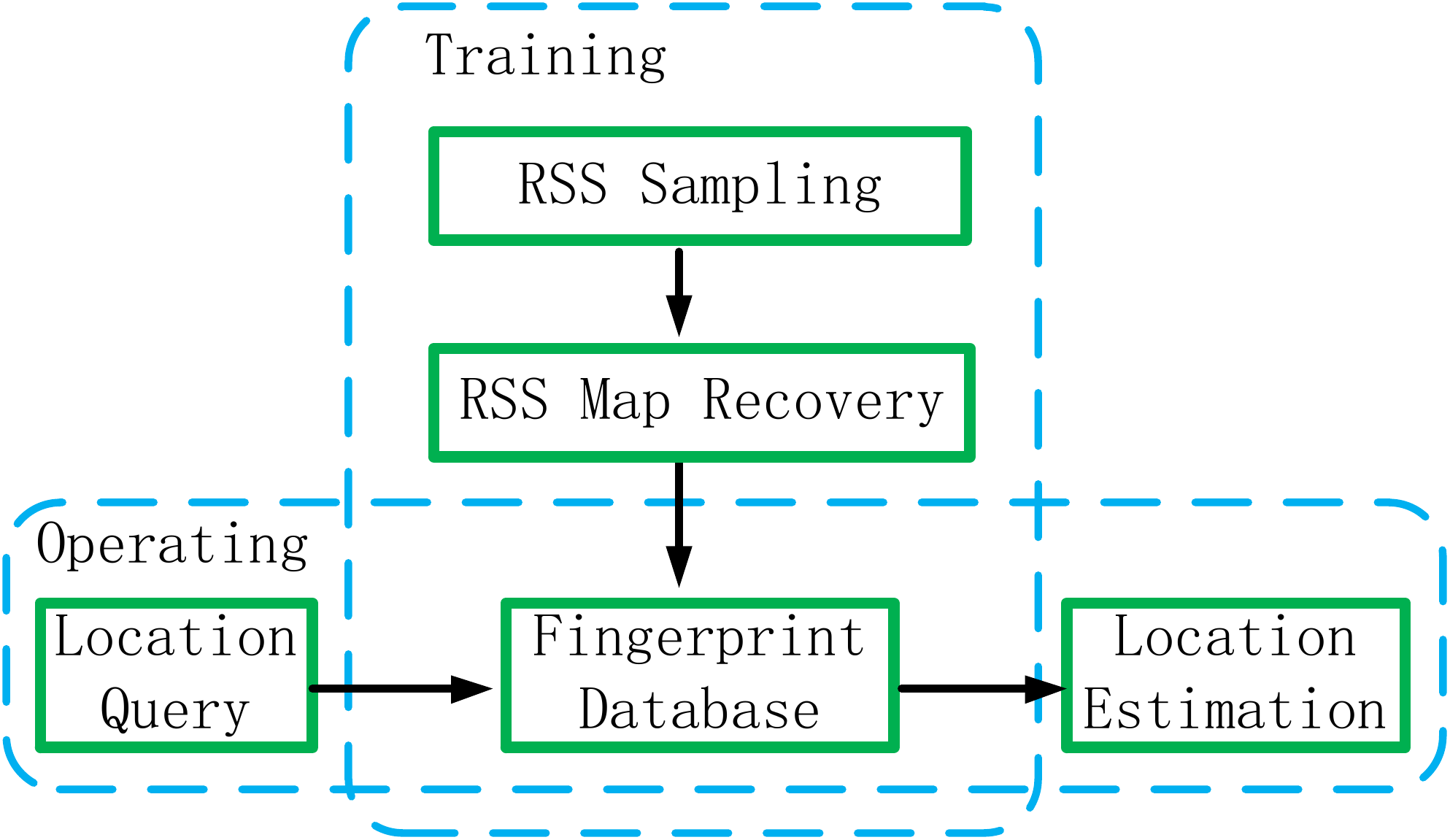}
\caption{Architecture of localization system.}
\label{fig2}
\end{figure}
\subsection{Indoor Localization System}
We present our indoor localization system architecture, as shown in Fig. \ref{fig2}. The fingerprint-based localization procedure can be divided into two phases: training and operating. The main task of training phase is constructing the fingerprint database. We first collect the fingerprint data at partial RPs, then recover the the fingerprints of other RPs based on the sampled data and construct a radio map, in which the mapping between RSS fingerprint and its corresponding location is stored. In the operating stage, a user sends a location query with current fingerprint, then the server matches the query fingerprint with constructed radio map to estimate his location and returns the result to the user.
\subsection{Overview of 3-D Tensors}
A third-order tensor is denoted by $\mathcal T\in \mathbb R^{N_1\times N_2\times N_3}$, and  its $(i,j,k)$-th entry is $\mathcal T(i,j,k)$. A tube (or fiber) of a tensor is a 1-D section defined by fixing all indices but one, which is a vector. A $slice$ of a tensor is a 2-D section defined by fix all but two indices. We will use $\mathcal T(:,:,k)$, $\mathcal T(:,j,:)$, and $\mathcal T(i,:,:)$ to denote the $frontal$, $lateral$ and $horizontal$ $slices$, respectively. In this paper, a vector $\mathcal T(i,j,:)$ denotes a fingerprint at reference point $(i,j)$.

Define $\breve{\mathcal T}(i,j,:)=\mathrm {fft}(\mathcal T(i,j,:))$, which denotes the Fourier transform along the third mode of $\mathcal T$. Similarly, one can obtain $\mathcal T$ from $\breve{\mathcal T}$ via inverse Fourier transform, i.e., $\mathcal T=\mathrm {ifft}(\breve{\mathcal T}, [\;], 3)$.

The $\mathrm {blkdiag}(\breve{\mathcal T})$ is a block diagonal matrix, which places the frontal slices in the diagonal, i.e., $\mathrm {diag}(\breve{\mathcal T}^{(1)}, \breve{\mathcal T}^{(2)}, ..., \breve{\mathcal T}^{(N_3)})$, where $\breve{\mathcal T}^{(i)}$ is the $i$th frontal slice of $\breve{\mathcal T}$, $i$=1, 2, ..., $N_3$.

The operation of a tensor-tensor product between two 3-D tensors is the foundation of algebraic development in [], which is defined as follows:

\noindent\textbf{Definition 1.} \textbf{\textit {t-product}}. \textit {The t-product $\mathcal K=\mathcal A*\mathcal B$ of $\mathcal A\in\mathbb R^{N_1\times N_2\times N_3}$ and $\mathcal B\in\mathbb R^{N_2\times N_4\times N_3}$ is a tensor of size $N_1\times N_4 \times N_3$ whose $(i,j)$th tube $\mathcal K(i,j,:)$ is given by $\mathcal K(i,j,:)=\sum_{k=1}^{N_2}\mathcal A(i,k,:)*\mathcal B(k,j,:)$, for} $i = $ \textrm{1, 2}, ..., $N_1$ \textit {and} $j = $ \textrm{1, 2} , ..., $N_4$.

One can view a third-order tensor of size $N_1\times N_2\times N_3$ as an $N_1\times N_2$ matrix of tubes. So the t-product means the multiplication of two matrices, and the circular convolution of two tubes replaces the multiplication of two elements.

\noindent\textbf{Definition 2.} \textbf{\textit {Tensor transpose}}. \textit {Let $\mathcal T$ be a tensor of size $N_1\times N_2\times N_3$, then $\mathcal T^\top$ is the $N_2\times N_1\times N_3$ tensor obtained by transposing each of the frontal slices and then reversing the order of transposed frontal slices 2 through $N_3$.}

\noindent\textbf{Definition 3.} \textbf{\textit {Identity tensor}}.  \textit {The identity tensor $\mathcal I\in \mathbb R^{N_1\times N_1\times N_3}$ is a tensor whose first frontal slice $\mathcal I(:,:,1)$ is the $N_1\times N_1$ identity matrix and all other frontal slices are zero}.

\noindent\textbf{Definition 4.} \textbf{\textit {Orthogonal tensor}}. \textit {A tensor $\mathcal Q\in\mathbb R^{N_1\times N_1\times N_3}$ is orthogonal if it satisfies $\mathcal Q^{\top}\ast \mathcal Q=\mathcal Q\ast \mathcal Q^{\top} =\mathcal I$}.

\noindent\textbf{Definition 5.} \textit {The inverse of a tensor $\mathcal V\in \mathbb R^{N_1\times N_1\times N_3}$ is written as $\mathcal V^{-1}\in \mathbb R^{N_1\times N_1\times N_3}$ and satisfies $\mathcal V^{-1}\ast \mathcal V=\mathcal V\ast \mathcal V^{-1}=\mathcal I$}.


\noindent\textbf{Definition 6.} \textbf{\textit {f-diagonal tensor}}. \textit {A tensor is called f-diagonal if each frontal slice of the tensor is a diagonal matirx}.

\noindent\textbf{Definition 7.} \textbf{\textit {t-SVD}}. \textit {For $\mathcal T\in\mathbb R^{N_1\times N_2\times N_3}$, the t-SVD of $\mathcal T$ is given by}
\begin{equation}
\mathcal T=\mathcal U*\Theta * \mathcal V^{\top},
\end{equation}
\textit {where $\mathcal U$ and $\mathcal V$ are orthogonal tensors of size $N_1\times N_1\times N_3$ and $N_2\times N_2\times N_3$ respectively. $\Theta$ is a rectangular f-diagonal tensor of size $N_1\times N_2\times N_3$.}

\noindent\textbf{Definition 8.} \textbf{\textit {Tensor tubal-rank}}. \textit {The tensor tubal-rank of a third-order tensor is the number of non-zero tubes of $\Theta$ in the t-SVD}.

\noindent\textbf{Definition 9.} \textit {The tensor-nuclear-norm (TNN) denoted by $\Vert \mathcal T \Vert_{TNN}$ and defined as the sum of the singular values of all the frontal slices $\breve{\mathcal T}$ is a norm and is the tightest convex relaxation to the $l_1$-norm of the tensor multi-rank. In fact, the tensor-nuclear-norm $\Vert \mathcal T \Vert_{TNN}$ is equivalent to the matrix nuclear-norm $\Vert \mathrm {blkdiag}(\breve{\mathcal T})\Vert_\ast$, $i.e.$, $\Vert \mathcal T \Vert_{TNN}=\Vert \mathrm {blkdiag}(\breve{\mathcal T})\Vert_\ast$}.
\subsection{System Model}
Suppose that the localization area is a rectangle $\mathcal R\in \mathbb R^2$. We divide the $\mathcal R$ into grids of same size. The grid map includes $N_1\times N_2$ grid points, which are called reference points (RPs). The grid map and the number of RP are denoted by $\mathcal C$ and $N=N_1\times N_2$, respectively. There are $N_3$ available access points (APs) in $\mathcal R$, which are randomly deployed and the locations are unknown. We use a third-order tensor $\mathcal T \in \mathbb R^{N_1\times N_2\times N_3}$ to denote the RSS map. For each RP $(i,j)$, the RSS fingerprint at the RP can be denoted as a vector $\mathcal T(i,j,:)$, where $\mathcal T(i,j,k)$ is the RSS value of the $k$th AP. Note that if the $k$th AP cannot be detected, we assume the noise level is equal to $-$110 dBm, namely, $\mathcal T(i,j,k)$=$-$110 dBm.

The radio map stores the RSS fingerprint at each RP and its corresponding coordinates. We use $\mathcal T(i,j,:)$ to denote the RSS map of $\mathcal C$, and the t-SVD is $\mathcal T = \mathcal U*\Theta * \mathcal V^{\top}$. The RSS value has high spatial correlation (columns and rows of $\mathcal C$) for each AP and also across these $N_3$ APs. Therefore we model this correlation by assuming that $\mathcal T$ has a low-rank tensor component $\mathcal X\in \mathbb R^{N_1\times N_2\times N_3}$, with tubal-rank $r$. Namely, there are only $r$ non-zero tubes in $\Theta$. On the other hand, the observed RSS values may be corrupted by anomaly readings. The anomalies are unknown a priori and sparsity. We assume that $\mathcal T$ has another sparse tensor component $\mathcal Y\in \mathbb R^{N_1\times N_2\times N_3}$. Therefore, we suppose the $\mathcal T$ is the superposition of the $\mathcal X$ and $\mathcal Y$. The model is defined as:
\begin{equation}
\mathcal T=\mathcal X+\mathcal Y.
\end{equation}
\subsection{Problem Statement}
It will be a time-consuming and labor-intensive work if we collect fingerprint data at all RPs. To reduce the workload in the site survey process, we sample fingerprints on a subset of RPs. Therefore, the site survey process is represented by using the following partial observation model under tubal-sampling:
\begin{equation}
\mathcal M=\mathcal P_\Omega(\mathcal X)+\mathcal P_\Omega(\mathcal Y),
\end{equation}
where $\mathcal M=\mathcal P_\Omega(\mathcal T)$ is an observation tensor, $\Omega$ being a subset of the grid map $\mathcal C$, the $(i,j,k)$th entry of $\mathcal P_\Omega(\mathcal S)$ is equal to $\mathcal S(i,j,k)$ if $(i,j)\in \Omega$ and zero otherwise. Note that the subset $\Omega$ can be obtained by non-adaptive or adaptive tubal-sampling approaches as proposed in \cite{liu2015fine}.

We focus on recovering a tensor $\mathcal X$ from a partial observation tensor $\mathcal M$. There are two properties about $\mathcal X$: the tensor $\mathcal X$ is low-tubal-rank, and the estimated $\hat{\mathcal X}$ should equal to $\mathcal M$ on the set $\Omega$. We can exploit the two properties to estimate $\mathcal X$ given the observation $\mathcal M$ by solving the following optimization problem:
\begin{eqnarray}
&&\text{Obj:} \langle \hat{\mathcal X}, \hat{\mathcal Y}\rangle = \arg\min\limits_{\langle \mathcal X, \mathcal Y\rangle}\;\; rank(\mathcal X)+\lambda\Vert \mathcal Y \Vert_{1,1,2}\nonumber\\
&&\text{s.t.}\quad \mathcal M= \mathcal P_\Omega(\mathcal X)+\mathcal P_\Omega(\mathcal Y).
\label{prob}
\end{eqnarray}
where $\mathcal X$ is the decision variable, rank($\cdot$) denotes the tensor tubal-rank, and $\lambda>0$ is a regularization parameter and the $\Vert \mathcal Y \Vert_{1,1,2}$ for 3D-tensor is defined as $\sum_{i,j}\Vert \mathcal Y(i,j,:) \Vert_F$ \cite{zhang2014novel}. The goal of the problem (\ref{prob}) is to estimate a tensor $\mathcal X$ with the least tubal-rank in given set $\Omega$ and observation $\mathcal M$ case.

Note that we must set a sufficient condition to guarantee unique solution for problem (\ref{prob}), which is defined as:
\begin{equation}
\Vert \mathcal P_{\Omega^c}(\mathcal Y) \Vert_F = 0.
\end{equation}
It is quite reasonable that our aim is to recover the low-rank tensor $\mathcal X$ and we do not want to recover the anomaly tensor \cite{LS2015}.
\section{Robust Tensor Completion}
We adopt uniform random tubal-sampling approach from the grid map $\mathcal C$ to produce observation. Given $\Omega$, the tensor tubal-rank function in (\ref{prob}) is combinatorial. Similar to the low-rank matrix completion case, the optimization problem (\ref{prob}) is NP-hard. In this case, the tubal-rank measure can be relaxed to a tensor nuclear norm (TNN) measure as studied in \cite{zhang2014novel}, which is applied in the problem of tensor completion and tensor principal component analysis.

Inspired by using the tensor-nuclear-norm for tensor completion in no anomaly case, we propose the following problem for the anomaly case, which is equivalent with (\ref{prob}) :
\begin{eqnarray}
&&\text{Obj:} \langle \hat{\mathcal X}, \hat{\mathcal Y}\rangle = \arg\min\limits_{\langle \mathcal X, \mathcal Y\rangle}\;\; \Vert \mathcal X \Vert_{TNN}+\lambda\Vert \mathcal Y \Vert_{1,1,2}\nonumber\\
&&\text{s.t.}\quad \mathcal M = \mathcal P_\Omega(\mathcal X)+\mathcal P_\Omega(\mathcal Y).
\label{equiva}
\end{eqnarray}
In order to solve the proposed convex optimization problem, we use the general ADMM. Based on the frame of ADMM, we introduce an intermediate variable $\mathcal Z$ and $\mathcal X=\mathcal Z$. Specifically, the augmented Lagrangian function of (\ref{equiva}) is
\begin{eqnarray}
\mathcal{L}(\mathcal X,\mathcal Z,\mathcal Y,\mathcal A,\mathcal B)&=&\Vert \mathcal Z \Vert_{TNN}+\lambda\Vert \mathcal Y \Vert_{1,1,2} \nonumber\\
&&+(\mathcal M(:)-\mathcal P_\Omega(\mathcal X(:)+\mathcal Y(:)))^T\mathcal A(:) \nonumber\\
&&+\frac{\mu}{2}\Vert \mathcal M-\mathcal P_\Omega(\mathcal X+\mathcal Y)\Vert_F^2 \nonumber\\
&&+(\mathcal X(:)-\mathcal Z(:))^T\mathcal B(:)\nonumber\\
&&+\frac{\rho}{2}\Vert \mathcal X-\mathcal Z\Vert_F^2.
\label{lag}
\end{eqnarray}
The ADMM generates the new iterate $(\mathcal X^{k+1}, \mathcal Z^{k+1}, \mathcal Y^{k+1}, \mathcal A^{k+1}, \mathcal B^{k+1})$ via the following specific form:
\begin{eqnarray}
\mathcal X^{k+1}&=&\arg\min\limits_{\mathcal X}\,(\mathcal M(:)-\mathcal P_\Omega(\mathcal X(:)+\mathcal Y(:)))^T\mathcal A(:) \nonumber\\
&&+\frac{\mu}{2}\Vert \mathcal M-\mathcal P_\Omega(\mathcal X+\mathcal Y)\Vert_F^2 \nonumber\\
&&+(\mathcal X(:)-\mathcal Z(:))^T\mathcal B(:) \nonumber\\
&&+\frac{\rho}{2}\Vert \mathcal X-\mathcal Z\Vert_F^2,
\end{eqnarray}
\begin{eqnarray}
\mathcal Z^{k+1}&=&\arg\min\limits_{\mathcal Z}\,\Vert \mathcal Z \Vert_{TNN}+(\mathcal X^{k+1}(:)-\mathcal Z(:))^T\mathcal B(:)\nonumber\\
&&+\frac{\rho}{2}\Vert \mathcal X^{k+1}-\mathcal Z\Vert_F^2,
\end{eqnarray}
\begin{eqnarray}
\mathcal Y^{k+1}&=&\arg\min\limits_{\mathcal Y}\,\lambda\Vert \mathcal Y \Vert_{1,1,2} \nonumber\\
&&+(\mathcal M(:)-\mathcal P_\Omega(\mathcal X^{k+1}(:)+\mathcal Y(:)))^T\mathcal A^k(:)\nonumber\\
&&+\frac{\mu}{2}\Vert \mathcal M-\mathcal P_\Omega(\mathcal X^{k+1}+\mathcal Y) \Vert_F^2,
\label{sy}
\end{eqnarray}
\begin{eqnarray}
\mathcal A^{k+1}&=&\mathcal A^k+\mu(\mathcal M-P_\Omega(\mathcal X^{k+1}+\mathcal Y^{k+1})),
\end{eqnarray}
\begin{eqnarray}
\mathcal B^{k+1}&=&\mathcal B^k+\rho(\mathcal X^{k+1}-\mathcal Z^{k+1})),
\end{eqnarray}
where $\mu$, $\rho$, $\mathcal A$ and $\mathcal B$ are multipliers, $\mathcal K(:)$ denotes vectorizing the tensor $\mathcal K$.

The solution for $\mathcal X^{k+1}$ is given by $\mathcal X^{k+1}=(\mathcal I+\mathcal P_\Omega)^{-1}(\mathcal Z^k+\mathcal M-P_\Omega(\mathcal Y^k+\frac{\mathcal A^k}{\mu})-\frac{\mathcal B^k}{\rho})$, where $\mathcal I$ is the identity tensor.

The solution for $\mathcal Z^{k+1}$ is given by:
\begin{eqnarray}
\mathcal Z^{k+1}&=&\arg\min\limits_{\mathcal Z}\,\Vert \mathcal Z \Vert_{TNN} \nonumber\\
&&+\frac{\rho}{2}\Vert \mathcal Z-(\mathcal X^{k+1}+\frac{\mathcal B^k}{\rho})\Vert_F^2.
\label{compx}
\end{eqnarray}
Let $\mathcal J=\mathcal X^{k+1}+\frac{\mathcal B^k}{\rho}$. By the definition, $\Vert \mathcal Z \Vert_{TNN}=\Vert \mathrm {blkdiag}(\breve{\mathcal Z})\Vert_\ast$, therefore, (\ref{compx}) is a standard nuclear norm minimization problem and can be solved by applying soft thresholding on the singular values of $\mathcal J$. Let $A=U\Sigma V^T$ be the singular value decomposition of $A$. We define the singular value soft-thresholding (shrinkage) operator as $\mathcal D_\varepsilon(A)=U\mathcal S_\varepsilon [\Sigma] V^T$, where $S_\varepsilon [\cdot]$ is an element-wise soft-thresholding operator defined as:
\begin{equation}
\setlength{\nulldelimiterspace}{0pt}
S_\varepsilon[x]=\left\{\begin{IEEEeqnarraybox}[\relax][c]{l's}
x-\varepsilon,&if  $x>\varepsilon,$\\
x+\varepsilon,&if  $x<-\varepsilon,$\\
0,& otherwise,
\end{IEEEeqnarraybox}\right.
\end{equation}
where $\varepsilon>0$. Therefore, the optimal solution to (\ref{compx}) is:
\begin{equation}
\mathrm {blkdiag}(\breve{\mathcal Z})=\mathcal D_\frac{1}{\mu}(\mathrm {blkdiag}(\mathrm {fft}(\mathcal J),[\;],3)),
\end{equation}
we can obtain $\breve{\mathcal Z}$ from $\mathrm {blkdiag}(\breve{\mathcal Z})$.

The solution for $\mathcal Y^{k+1}$ is given by:
\begin{eqnarray}
\mathcal Y^{k+1}&=&\arg\min\limits_{\mathcal Y}\,\lambda\Vert \mathcal Y \Vert_{1,1,2} \nonumber\\
&&+\frac{\mu}{2}\Vert \mathcal M-\mathcal P_\Omega(\mathcal X^{k+1}+\mathcal Y)+\frac{\mathcal A^k}{\mu} \Vert_F^2,
\end{eqnarray}
so $\mathcal Y$ is given by:
\begin{equation}
\mathcal Y^{k+1}(i,j,:)=(1-\frac{\lambda}{\mu\Vert \mathcal Y^k(i,j,:)\Vert_F})_{+}\mathcal Y^k(i,j,:).
\end{equation}

For algorithm implementation, one can initialize $\mathcal A^0$ and $\mathcal B^0$ as zero tensors. $\Omega$ is an indicator matrix of size $N_1\times N_2$. The iterative algorithm for (\ref{equiva}) is proposed in Algorithm 1.

\begin{algorithm}[htb] 
\renewcommand{\algorithmicrequire}{\textbf{Input:}}
\renewcommand\algorithmicensure {\textbf{Output:} }
\caption{ Robust Tensor Completion } 
\label{alg:1} 
\begin{algorithmic}[1] 
\REQUIRE  
parameters $\Omega$, $M$, $N_1$, $N_2$, $N_3$, $\lambda$, $\mu$, $\rho$, $\mathcal A^0$, $\mathcal B^0$.
\STATE Initialize: $\mathcal A^0\leftarrow 0$; $\mathcal B^0\leftarrow 0$; $k\leftarrow 0$
\STATE \textbf{while} not converged \textbf{do}
\STATE $\mathcal X^{k+1}=(\mathcal I+\mathcal P_\Omega)^{-1}(\mathcal Z^k+\mathcal M-P_\Omega(\mathcal Y^k+\frac{\mathcal A^k}{\mu})-\frac{\mathcal B^k}{\rho})$
\STATE $\mathrm {blkdiag}(\breve{\mathcal Z})=\mathcal D_\frac{1}{\mu}(\mathrm {blkdiag}(\mathrm {fft}(\mathcal J),[\;],3))$
\STATE $\mathcal Y^{k+1}(i,j,:)=(1-\frac{\lambda}{\mu\Vert \mathcal Y^k(i,j,:)\Vert_F})_{+}\mathcal Y^k(i,j,:)$
\STATE $\mathcal A^{k+1}=\mathcal A^k+\mu(\mathcal M-P_\Omega(\mathcal X^{k+1}+\mathcal Y^{k+1}))$
\STATE $\mathcal B^{k+1}=\mathcal B^k+\rho(\mathcal X^{k+1}-\mathcal Z^{k+1}))$
\STATE $k\leftarrow k+1$
\STATE \textbf{end while}
\ENSURE  
$\mathcal X$, $\mathcal Y$
\end{algorithmic}
\end{algorithm}
\section{Performance Evaluation}
In this section, we present the evaluation methodology, which includes these aspects: ground truth, compared approaches and metric. Then the experiment results show the fingerprint database recovery performance and localization performance of the proposed and compared approaches.
\subsection{Methodology}
\begin{figure}[H]
\centering
\includegraphics[width=2in]{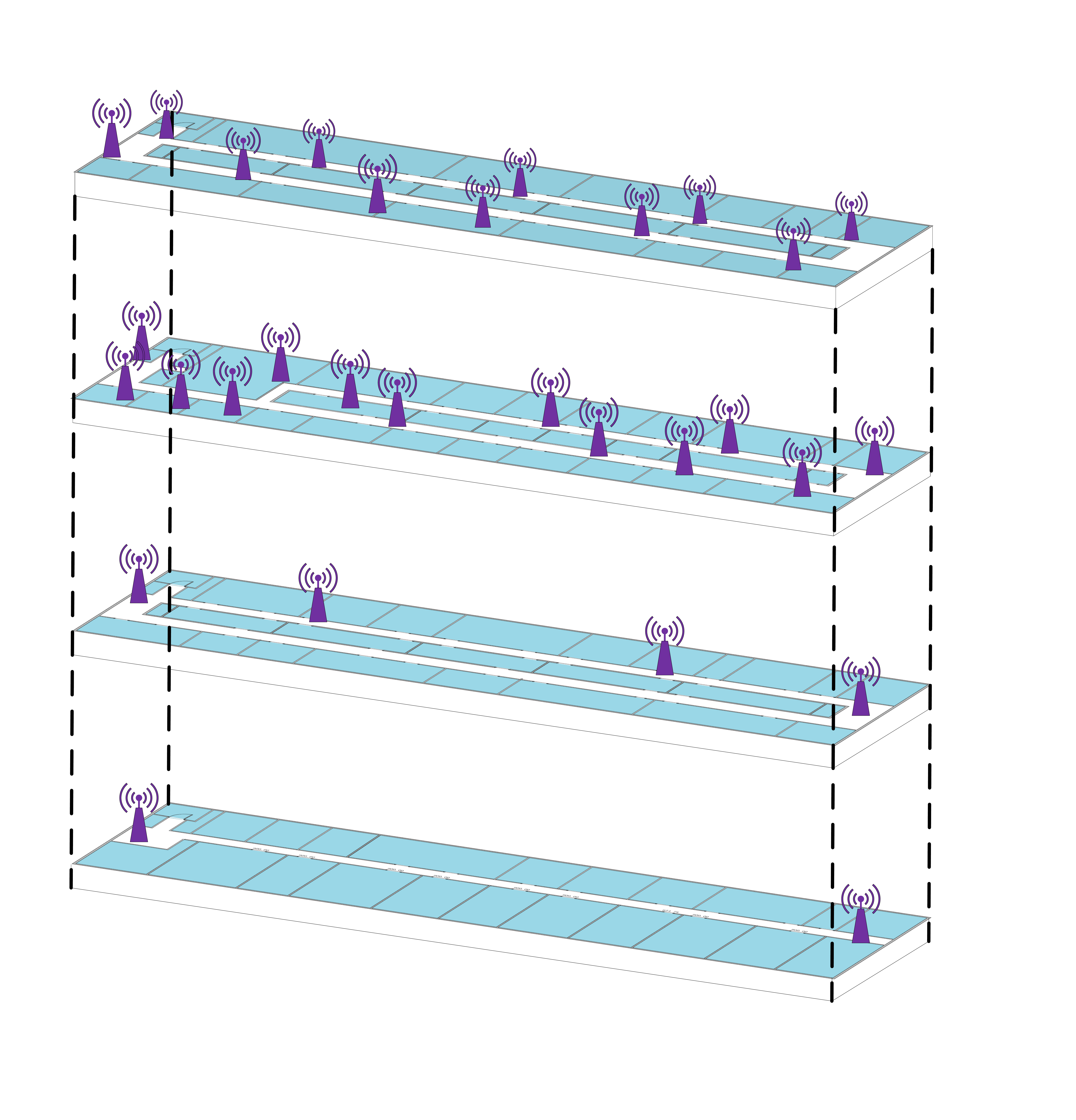}
\caption{SIMT-Net.}
\label{fig3}
\end{figure}

The indoor localization system is named as SIMT-Net, as shown in Fig. \ref{fig3}. Our experiment field of 80\,m $\times$ 20\,m is on the fourth floor of a real office building. It is divided into 268 $\times$ 63 grid map. There are 30 APs deployed in the whole building, which can be detected in the experiment field of the fourth floor. We collect the fingerprints at each RP in the grid map and use them to build a full third-order tensor as the ground truth. To be specific, for each RP, we set the average of RSS time samples as its RSS vector. The ground truth tensor has dimension 268 $\times$ 63 $\times$ 30. Note that the RSS values are measured in dBm.

To verify the effectiveness of the proposed approach, namely, tensor completion with anomaly readings (TCwA), two approaches for data recovery are chosen for comparison. One of them is the standard tensor completion (STC) approach  with not considering anomaly readings. Another is the oracle solution (OS) served as the optimal solution \cite{5513535}.

We focus on two kinds of performance: recovery error and localization error. To recover the complete measurements in the test area, we randomly sample a subset of all RPs as the known measurements. Varying of the sampling rate is from 10$\%$ to 90$\%$. We quantify the recovery error in terms of normalized square of error (NSE) for entries which are not included in the samples. The NSE is defined as:
\begin{equation}
NSE=\frac{\Sigma_{(i,j)\in\Omega^c}\Vert \hat{\mathcal T}(i,j,:)-\mathcal T(i,j,:)\Vert_F^2}{\Sigma_{(i,j)\in\Omega^c}\Vert \mathcal T(i,j,:)\Vert_F^2},
\end{equation}
where $\hat{\mathcal T}$ is the estimated tensor, $\Omega^c$ is the completion of set $\Omega$.

To further evaluate the efficiency and robust of the constructed radio map, the estimated tensor $\hat{\mathcal T}$ is used for localization. Here, we adopt classical localization technique, K-nearest neighbor (KNN), in the operating phase. In our experiments, we uniformly select 200 testing points within localization area and then exploit KNN algorithm to perform location estimation. The localization error is defined as the Euclidean distance between the estimated location and the real location of the testing point.
\subsection{Fingerprint Database Recovery Performance}
\begin{figure}[H]
\centering
\includegraphics[width=3in]{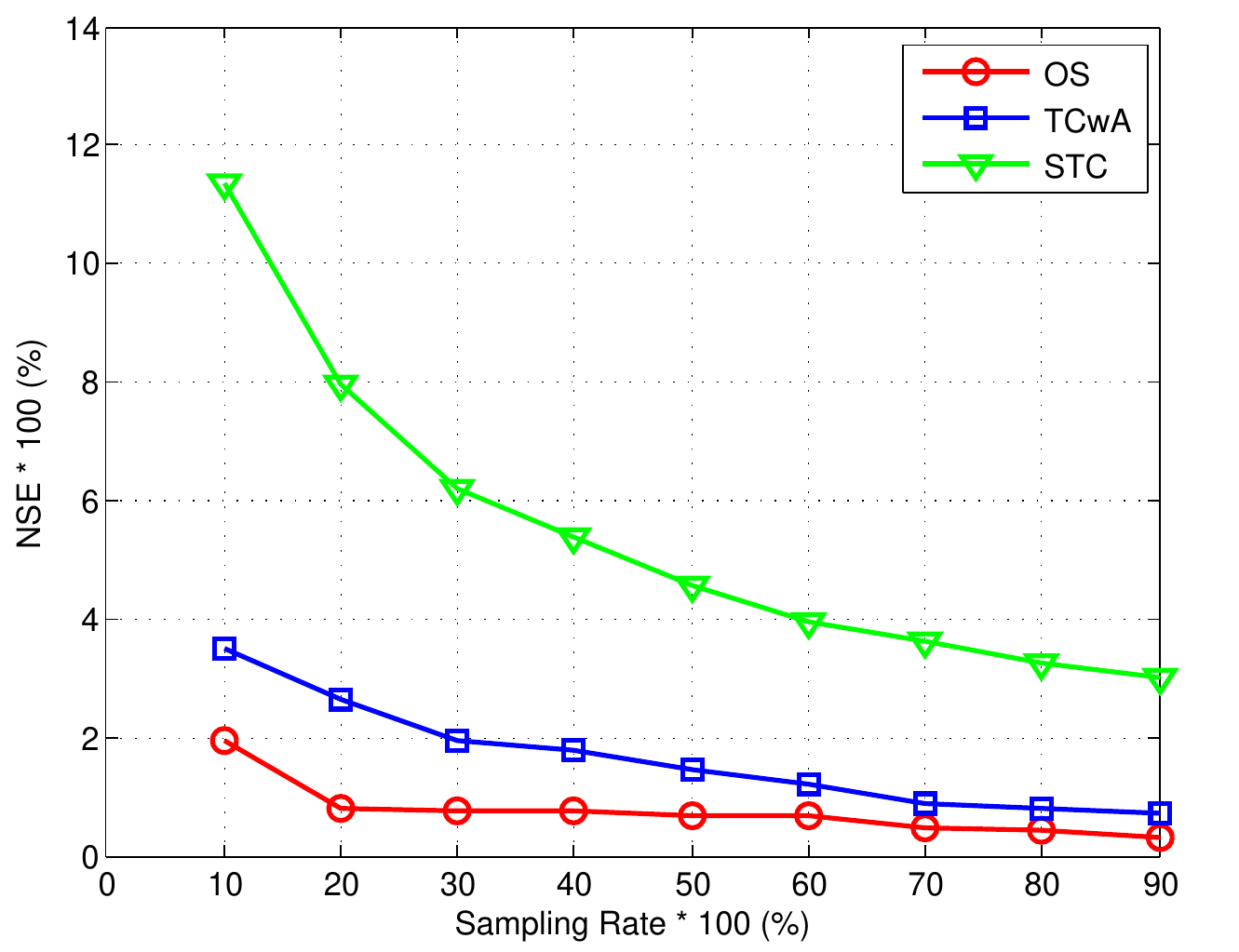}
\caption{Comparison of different approaches.}
\label{fig4}
\end{figure}

Figure. \ref{fig4} shows the RSS tensor recovery performance for varying sampling rate. Experimental results show the recovery errors for these three algorithms decrease with the sampling rate increases. It can be found that TCwA yields better performance than STC which do not take anomaly into consideration. Concretely, our TCwA approach achieves recovery error $\leq$ 4$\%$ for sampling rate $\geq $10$\%$, while STC approach has error 4$\%$ at the sampling rate 60$\%$. The result indicates that TCwA has robust behavior.
\begin{figure}[H]
\centering
\includegraphics[width=3in]{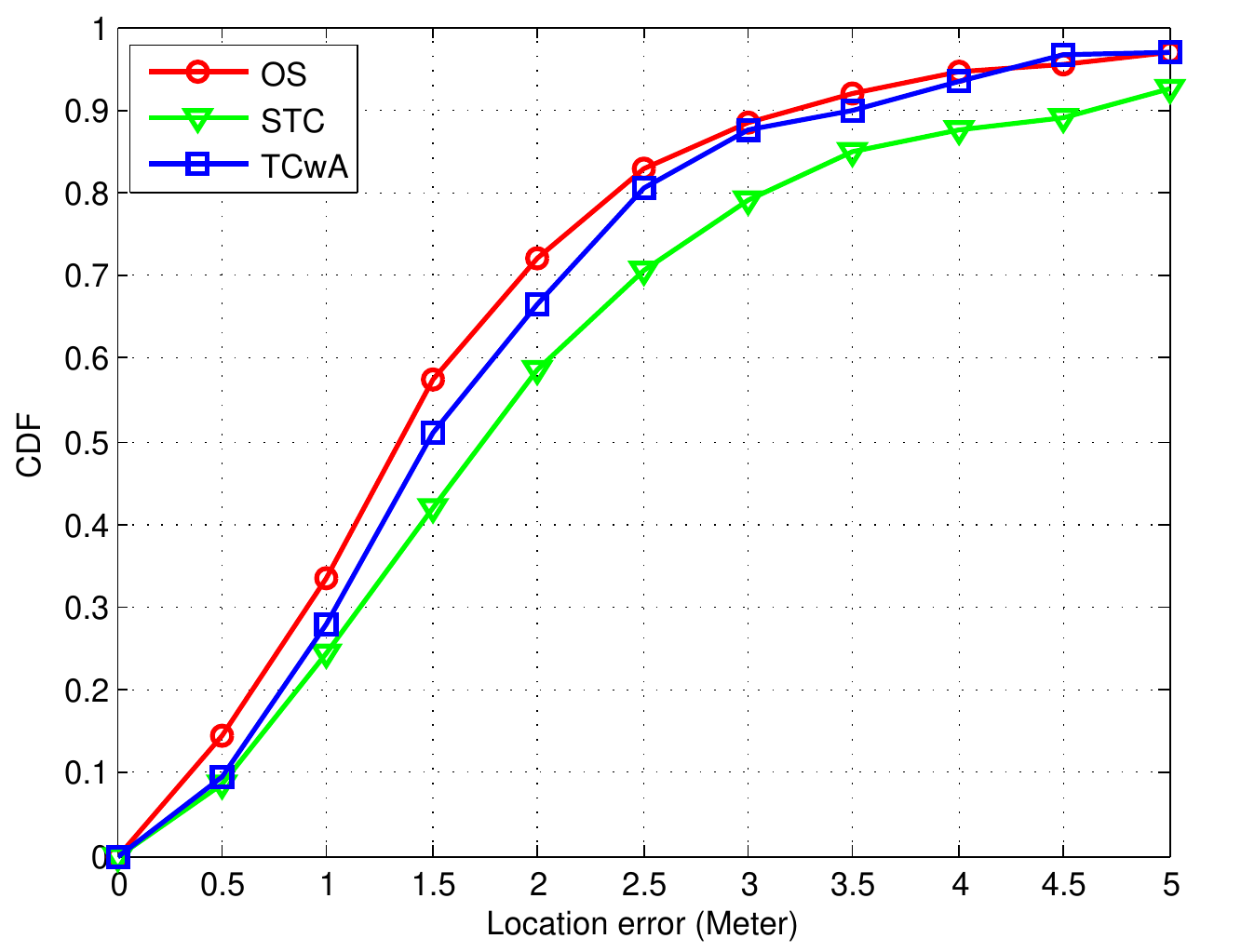}
\caption{The CDFs of localization error of KNN with 20\% sampling rate.}
\label{fig5}
\end{figure}
\subsection{Localization Performance}
The anomaly readings have large influences on the radio map construction and localization error. To further evaluate the efficiency and robustness of the constructed radio map, the constructed results are used for location estimation.

In our experiment, parts of RSS fingerprints of the RPs are used to construct fingerprint database while the rest are served as test. We select randomly fingerprints of 200 RPs as the test data from the rest. Fig. \ref{fig5} shows the empirical CDFs of localization error of KNN with 20$\%$ sampling rate. Our approach TCwA yields better performance compared with the STC approach. For example, the TCwA achieves 2.5\,m (80$\%$) localization error, while STC has about 3.1\,m (80$\%$) localization error. The difference of localization accuracy between TCwA and OS is small. It is shown that the localization results are consistent with the fingerprint database construction results.
\section{Conclusions}
In this paper, we study the problem of constructing fingerprint database in the site survey process for indoor localization. We propose a tensor decomposition model in the anomaly case for fingerprint data recovery. According to the low-tubal-rank feature of tensor and relationship between tensor-nuclear-norm and matrix nuclear-norm, we exploit ADMM algorithm to perform tensor completion and reconstruct the fingerprint database. Experiment results show that the proposed approach significantly reduces the number of measurements for fingerprint database construction while guaranteeing a higher recovery accuracy of fingerprint data. It behaves robustly to anomaly readings interference. Furthermore,  the localization accuracy is also enhanced compared to  the existing approach.
\bibliographystyle{IEEEtran}
\bibliography{IEEEabrv,tensor_ref}
\end{document}